\documentclass[aps,prl,showpacs,twocolumn,floats,epsfig]{revtex4}
\usepackage{amssymb}
\usepackage{amsbsy}
\usepackage{amsmath}
\usepackage{epsfig}
\newcommand\beq{\begin{equation}}
\newcommand\eeq{\end{equation}}
\newcommand\bea{\begin{eqnarray}}
\newcommand\eea{\end{eqnarray}}
\newcommand\non{\nonumber}

\begin{document}

\title{Tuning the conductance of Dirac fermions on the surface of a topological
insulator}

\author{S. Mondal,$^1$ D. Sen,$^2$ K. Sengupta,$^1$ and R. Shankar$^3$}

\affiliation{$^1$Theoretical Physics Division, Indian Association for the
Cultivation of Sciences, Kolkata 700 032, India \\
$^2$ Center for High Energy Physics, Indian Institute of Science, Bangalore
560 012, India \\
$^3$ The Institute of Mathematical Sciences, C.I.T Campus, Chennai 600 113,
India}

\date{\today}

\begin{abstract}
We study the transport properties of the Dirac fermions with Fermi
velocity $v_F$ on the surface of a topological insulator across a
ferromagnetic strip providing an exchange field ${\mathcal J}$ over
a region of width $d$. We show that the conductance of such a
junction changes from oscillatory to a monotonically decreasing
function of $d$ beyond a critical ${\mathcal J}$. This leads to the
possible realization of a magnetic switch using these junctions. We
also study the conductance of these Dirac fermions across a
potential barrier of width $d$ and potential $V_0$ in the presence
of such a ferromagnetic strip and show that beyond a critical
${\mathcal J}$, the criteria of conductance maxima changes from
$\chi= e V_0 d/\hbar v_F = n \pi$ to $\chi= (n+1/2)\pi$ for integer
$n$. We point out that these novel phenomena have no analogs in
graphene and suggest experiments which can probe them.
\end{abstract}


\maketitle

Topological insulators in both two- and three- dimensions (2D and
3D) have attracted a lot of theoretical and experimental attention
in recent years \cite{zhang1,hassan1,kane1,kane2}. It has been shown
in Ref.\ \cite{kane2} that such 3D insulators can be completely
characterized by four integers $\nu_0$ and $\nu_{1,2,3}$. The former
specifies the class of topological insulators to be strong
($\nu_0=1$) or weak ($\nu_0=0$), while the latter integers
characterize the time-reversal invariant momenta of the system given
by $\vec M_0 = (\nu_1 \vec b_1, \nu_2 \vec b_2, \nu_3 \vec b_3)/2$,
where $\vec b_{1,2,3}$ are the reciprocal lattice vectors. The
topological features of strong topological insulators (STI) are
robust against the presence of time-reversal invariant perturbations
such as disorder or lattice imperfections. It has been theoretically
predicted \cite{kane2,zhang1} and experimentally verified
\cite{hassan1} that the surface of a STI has an odd number of Dirac
cones whose positions are determined by the projection of $\vec M_0$
on to the surface Brillouin zone. The position and number of these
cones depend on both the nature of the surface concerned and the
integers $\nu_{1,2,3}$. For several compounds such as $\rm HgTe$ and
${\rm Bi_2 Se_3}$, specific surfaces with a single Dirac cone near
the $\Gamma$ point of the 2D Brillouin zone have been found
\cite{hassan1,exp2}. Such a Dirac cone is described by the
Hamiltonian \bea H ~=~ \int \frac{dk_x dk_y}{(2\pi)^2}
~\psi^{\dagger}(\vec k) ~(\hbar v_F {\vec \sigma} \cdot {\vec k} -
\mu I) ~\psi(\vec k), \label{ham1} \eea where $\vec \sigma (I)$
denotes the Pauli (identity) matrices in spin space, $\psi=
(\psi_{\uparrow}, \psi_{\downarrow})^T$ is the annihilation operator
for the Dirac spinor, $v_F$ is the Fermi velocity, and $\mu$ is the
chemical potential \cite{kane4}. Recently, several novel features of
these surface Dirac electrons such as the existence of Majorana
fermions in the presence of a magnet-superconductor interface on the
surface \cite{kane4,been1,tanaka2}, generation of a time-reversal
symmetric $p_x+ip_y$ superconducting state via proximity to a
$s$-wave superconductor \cite{kane4}, anomalous magnetoresistance of
ferromagnet-ferromagnet junctions \cite{tanaka1} and novel spin
textures with chiral properties \cite{hassan2} have been studied in
detail.

\begin{figure} \includegraphics[width=\linewidth]{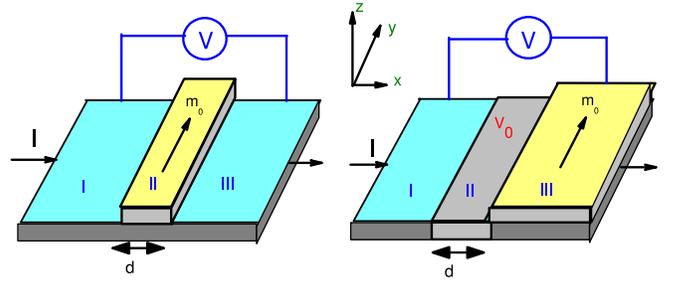}
\caption{Proposed experimental setups: a) Left panel: The
ferromagnetic film extends over region II of width $d$ providing an
exchange field in this region. b) Right panel: The film extends over
region III while the region II has a barrier characterized by a voltage
$V_0$. $V$ and $I$ denote the bias voltage and current across the junction
respectively. See text for details.} \label{fig1} \end{figure}

In this letter, we study the transport properties of these surface
Dirac fermions in two experimentally realizable situations shown in
Fig.\ \ref{fig1}. The first study concerns their transport across a
region with a width $d$ with a proximity-induced exchange field
${\mathcal J}$ arising from the magnetization $\vec m = m_0 \hat y$
of a proximate ferromagnetic film as shown in the left panel of
Fig.\ \ref{fig1}. We demonstrate that the tunneling conductance $G$
of these Dirac fermions through such a junction can either be an
oscillatory or a monotonically decaying function of the junction
width $d$. One can interpolate between these two qualitatively
different behaviors of $G$ by changing $m_0$ (and thus ${\mathcal
J}$) by an applied in-plane magnetic field leading to the possible
use of this junction as a magnetic switch. The second study concerns
the transport properties of Dirac fermions across a barrier
characterized by a width $d$ and a potential $V_0$ in region II with
a magnetic film proximate to region III as shown in the right panel
of Fig.\ \ref{fig1}. We note that it is well known from the context
of Dirac fermions in graphene \cite{neto1} that such a junction, in
the absence of the induced magnetization, exhibits transmission
resonances with maxima of transmission at $\chi= e V_0 d/\hbar v_F =
n \pi$, where $n$ is an integer. Here we show that beyond a critical
strength of $m_0$, the maxima of the transmission shifts to $\chi=
(n+1/2) \pi$. Upon further increasing $m_0$, one can reach a regime
where the conductance across the junctions vanishes. We stress that
the properties of Dirac fermions elucidated in both these studies
are a consequence of their spinor structure in physical spin space,
and thus have no analogs for either conventional Schr\"odinger
electrons in 2D or Dirac electrons in graphene \cite{kat1}.

We begin with an analysis of the junction shown in the left panel of
Fig.\ \ref{fig1}. The Dirac fermions in region I and III are
described by the Hamiltonian in Eq. (\ref{ham1}). Consequently, the
wave functions of these fermions moving along $\pm x$ in these
regions for a fixed transverse momentum $k_y$ and energy $\epsilon$
can be written as \bea \psi_i^{\pm} ~=~ (1, \pm e^{\pm i \alpha})
~e^{i (\pm k_x x + k_y y)}/ \sqrt{2}, \label{nwave1} \eea where $i$
takes values I and III, $\alpha = \arcsin(\hbar v_F k_y/|\epsilon
+\mu|)$ and $k_x(\epsilon)= \sqrt{[(\epsilon+\mu)/\hbar v_F]^2 -
k_y^2}$. In region II, the presence of the ferromagnetic strip with
a magnetization $\vec m_0 = m_0 \hat y$ leads to the additional term
$ H_{\rm induced} = \int dx dy \, {\mathcal J} \theta(x) \theta(d-x)
\psi^{\dagger} (\vec x) \sigma_y\, \psi(\vec x)$, where ${\mathcal
J} \sim m_0$ is the exchange field due to the presence of the strip
\cite{tanaka1}, and $\theta(x)$ denotes the Heaviside step function.
Note that $H_{\rm induced}$ may be thought as a vector potential
term arising due to a fictitious magnetic field $\vec B_{f}=
({\mathcal J}/ev_F) [\delta(x)-\delta(d-x)] \hat z$. This analogy
shows that our choice of the in-pane magnetization along $\hat y$ is
completely general; all gauge invariant quantities such as
transmission are independent of $x$-component of $\vec m_0$ in the
present geometry. For a given $m_0$, the precise magnitude of
${\mathcal J}$ depends on several factors such as the exchange
coupling of the film and can be tuned, for soft ferromagnetic films,
by an externally applied field \cite{tanaka1}. The wave function for
the Dirac fermions in region II moving along $\pm x$ in the presence
of such an exchange field is given by \bea \psi_{II}^{\pm} ~=~ (1,
\pm e^{\pm i \beta}) ~e^{i( \pm k_x^{'} x+ k_y y)}/ \sqrt{2},
\label{magwave1} \eea where $\beta = \arcsin(\hbar v_F (k_y +
M)/|\epsilon +\mu|)$, $M= {\mathcal J}/(\hbar v_F)$, and
$k_x^{'}(\epsilon)= \sqrt{[(\epsilon+\mu)/ \hbar v_F]^2 -
(k_y+M)^2}$. Note that beyond a critical $M_c= \pm
2|\epsilon+\mu|/(\hbar v_F)$ (and hence a critical ${\mathcal J}_c =
\pm 2 |\epsilon+\mu|$), $k^{'}_x$ becomes imaginary for all $k_y$
leading to spatially decaying modes in region II.

Let us now consider an electron incident on region II from the left
with a transverse momentum $k_y$ and energy $\epsilon$. Taking into
account reflection and transmission processes at $x=0$ and $x=d$,
the wave function of the electron can be written as $\psi_I =
\psi_I^{+} + r \psi_I^{-}$, $\psi^{II} = p \psi_{II}^{+} + q
\psi_{II}^{-}$, and $\psi_{III} = t \psi_{III}^{+}$. Here $r$ and
$t$ are the reflection and transmission amplitudes and $p$ ($q$)
denotes the amplitude of right (left) moving electrons in region II.
Matching boundary conditions on $\psi_I$ and $\psi_{II}$ at $x=0$
and $\psi_{II}$ and $\psi_{III}$ at $x=d$ leads to
\bea 1+r &=& p+q, \quad e^{i \alpha} -r e^{-i\alpha} = p e^{i \beta} - q
e^{-i \beta}, \non \\
te^{i k_x d}&=& p e^{ik_x^{'}d} + q e^{-ik_x^{'}d}, \non \\
te^{i (k_x d+\alpha)}&=& p e^{i(k_x^{'}d +\beta)} - q e^{-i(k_x^{'}d+\beta)}.
\label{cond1} \eea

Solving for $t$ from Eq. (\ref{cond1}), one finally obtains the
conductance $G = dI/dV = (G_0/2) \int_{-\pi/2}^{\pi/2} T \,
\cos(\alpha) d \alpha$. Here $G_0 = \rho(eV) w e^2/(\pi \hbar^2 v_F)
$, $\rho(eV)= |(\mu+eV)|/ [2\pi (\hbar v_F)^2]$ is the density of
states (DOS) of the Dirac fermions and is a constant for $\mu \gg eV$,
$w$ is the sample width, and the transmission $T = |t|^2$ is given by
\bea T &=& \cos^2(\alpha) \cos^2(\beta)/[\cos^2(k_x^{'} d)\cos^2(\alpha)
\cos^2(\beta) \non \\
&& + \sin^2(k_x^{'} d) ( 1-\sin(\alpha) \sin(\beta))^2]. \label{trans1} \eea

\begin{figure} \includegraphics[width=0.95\linewidth]{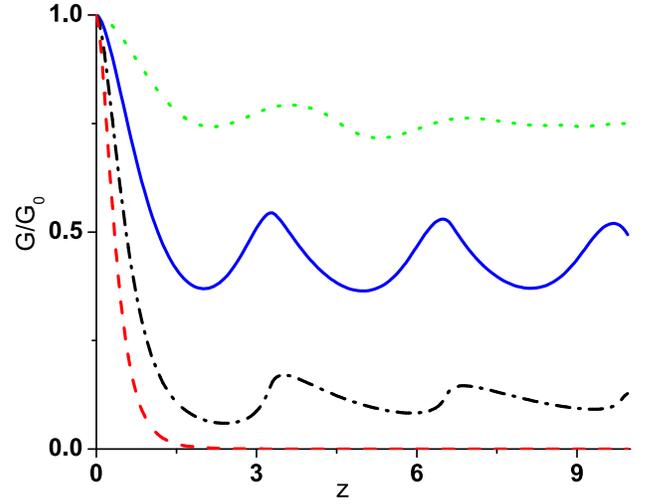}
\caption{Plot of tunneling conductance $G/G_0$ for a fixed $V$ and
$\mu$ as a function of the effective width $z=d|eV+\mu|/\hbar v_F$
for $\hbar v_F M/|eV+\mu| =0.3$ (green dotted line), $0.7$ (blue solid
line), $1.3$ (black dash-dotted line) and $2.1$ (red dashed line).
The value of the critical $M$ is given by $\hbar v_F M/|eV+\mu|=2$. See
text for details.} \label{fig2} \end{figure}

\begin{figure} \includegraphics[width=0.9 \linewidth]{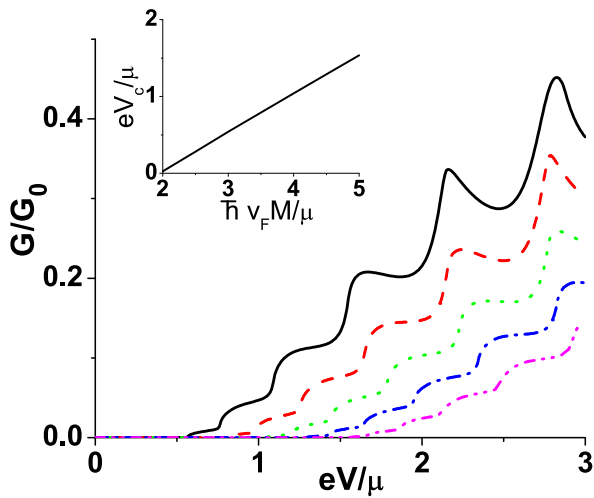}
\caption{Plot $G/G_0$ versus $eV/\mu$ for several
representatives values $\hbar v_F M/\mu$ ranging from $3$ (left-most
black solid curve) to $5$ (right-most magenta dash-double dotted
line) in steps of $0.5$ The effective junction width $z_0=5$ for all
plots. The inset shows a plot of $eV_c/\mu$ versus $\hbar
v_F M/\mu$. See text for details.} \label{fig3} \end{figure}

Eq. (\ref{trans1}) and the expression for $G$ represent one of the
main results of this work. We note that for a given $\alpha$, $T$
has an oscillatory (monotonically decaying) dependence on $d$
provided $k_x^{'}$ is real (imaginary). Since $k_x^{'}$ depends, for
a given $\alpha$, on $M$, we find that one can switch from an
oscillatory to a monotonically decaying $d$ dependence of
transmission in a given channel (labeled by $k_y$ or equivalently
$\alpha$) by turning on a magnetic field which controls $m_0$ and
hence $M$. Also since $-1 \le \sin(\alpha)\le 1$, we find that
beyond a critical $M=M_c$, the transmission in all channels exhibits
a monotonically decaying dependence on $d$. Consequently, for a
thick enough junction one can tune $G$ at fixed $V$ and $\mu$ from a
finite value to nearly zero by tuning $M$ ({\it i.e.}, $m_0$)
through $M_c$. Thus such a junction may be used as a magnetic
switch. These qualitatively different behaviors of the junction
conductance $G$ for $M$ below and above $M_c$ is demonstrated in
Fig.\ \ref{fig2} by plotting $G$ as a function of effective barrier
width $z=d|eV+\mu|/\hbar v_F$ for several representative values of
$\hbar v_F M/|eV+\mu|$. Since $T$ and hence $G$ depends on $M$
through the dimensionless parameter $\hbar v_F M/|eV+\mu|$, this
effect can also be observed by varying the applied voltage $V$ for a
fixed $\mu$, $d$, and $M$. In that case, for a reasonably large
dimensionless barrier thickness $z_0=d \mu/\hbar v_F$, $G/G_0$
becomes finite only beyond a critical voltage $|eV_c + \mu| = \hbar
v_F M/2$ as shown in Fig.\ \ref{fig3} for several representative
values of $z_0$. This critical voltage $V_c$ can be determined
numerically by finding the lowest voltage for which $G/G_0$ exhibits
a monotonic decay as a function of $z_0$. The plot of $eV_c/\mu$ as
a function of $\hbar v_F M/\mu$, shown in inset of Fig.\ \ref{fig3},
demonstrates the expected linear relationship between $V_c$ and $M$.
We note such a magnetic field or applied bias voltage dependence of
the junction conductance necessitates that the Dirac electrons
represents spinors in physical spin space and is therefore
impossible to achieve in graphene \cite{neto1}.

Next, we analyze the junction shown in the right panel of Fig.\
\ref{fig1} where the region III below a ferromagnetic film is
separated from region I by a potential barrier in region II. Such a
barrier can be applied by changing the chemical region in region II
either by a gate voltage $V_0$ or via doping \cite{exp2}. In the
rest of this work, we will analyze the problem in the thin barrier
limit for which $V_0 \to \infty$ and $d\to 0$, keeping the
dimensionless barrier strength $\chi= e V_0 d/(\hbar v_F)$ finite.
The wave function of the Dirac fermions moving along $\pm x$ with a
fixed momentum $k_y$ and energy $\epsilon$ in this region is given
by \beq \psi^{'\,\pm}_{II} ~=~ (1, \pm e^{\pm i \gamma}) ~e^{i (\pm
k^{''}_x x + k_y y)}/\sqrt{2}, \label{barwave1} \eeq where $\gamma
=\arcsin( \hbar v_F k_y/|\epsilon + eV_0+\mu|)$ and $k^{''}_x
(\epsilon) = \sqrt{[(\epsilon+eV_0+\mu)/\hbar v_F]^2 - k_y^2}$. The
wave functions in region I and III are given by Eqs. (\ref{nwave1})
and (\ref{magwave1}) respectively: $\psi^{'}_I = \psi_I$ and
$\psi^{'}_{III} = \psi_{II}$. Note that one can have a propagating
solution in region III only if $|M| \le |M_c|$.

\begin{figure} \includegraphics[width=0.95 \linewidth]{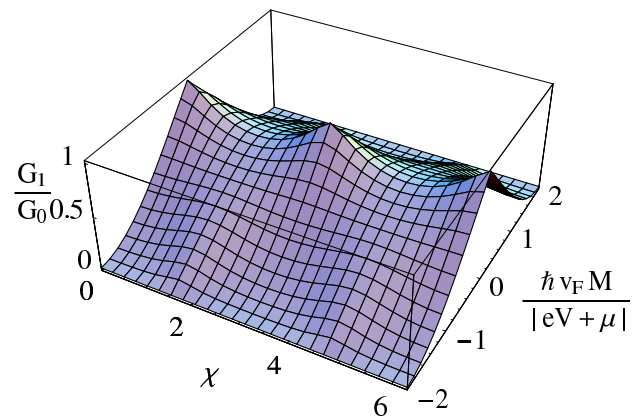}
\caption{Plot of tunneling conductance $G_1/G_0$ versus
the effective barrier strength $\chi$ and $\hbar v_F M$ for fixed
applied voltage $V$ and chemical potential $\mu$. $G_1$ vanishes for
$|M| \ge M_c=2|eV+\mu|/\hbar v_F$.} \label{fig4} \end{figure}

The transmission problem for such a junction can be solved by an
procedure similar to the one outlined above for the magnetic strip
problem. For an electron approaching the barrier region from the
left, we write down forms of the wave function in the three regions
I, II and III: $\psi^{'}_I = \psi_I^{+} + r_1 \psi_I^{-}$,
$\psi^{'}_{II} = p_1 \psi^{'\,+}_{II} + q_1 \psi^{'\,-}_{II}$, and
$\psi^{'}_{III} = t_1 \psi_{II}^{+}$. As outlined earlier, one can
then match boundary conditions at $x=0$ and $x=d$, and obtain the
transmission coefficient $T_1=|t_1|^2 k^{'}_x/k_x$ as
\bea T_1 &=& 2 \cos(\beta) \cos(\alpha)/[1+\cos(\beta-\alpha) \non \\
&& -\cos^2(\chi)\{\cos(\beta -\alpha)-\cos(\beta +\alpha)\}].
\label{trans2} \eea
Note that in the absence of the ferromagnetic
film over region III, $\beta=\alpha$, and $T_1 \to T^0_1 =
\cos^2(\alpha)/[1-\cos^2(\chi) \sin^2(\alpha)]$. The expression for
$T_1^0$, reproduced here for the special case of $M=0$, is
well known from analogous studies in the context of graphene, and it
exhibits both Klein paradox ($T^0_1=1$ for $\alpha=0$) and
transmission resonance ($T^0_1=1$ for $\chi=n \pi $) \cite{kat1}.
When $M\ne 0$, we find that the transmission for normal incidence
($k_y=0$) does become independent of the barrier strength, but its
magnitude deviates from unity: $T^{\rm normal}_1 = 2\sqrt{1-(\hbar
v_F M/|eV +\mu|)^2}/(1+ \sqrt{1-(\hbar v_F M/|eV +\mu|)^2})$. The
value of $T^{\rm normal}_1$ decreases monotonically from $1$ for
$M=0$ to $0$ for $|M| = |eV +\mu|/(\hbar v_F)$ and can thus be tuned
by changing $M$ (or $V$) for a fixed $V$ (or $M$) and $\mu$.

\begin{figure} \includegraphics[width=0.9 \linewidth]{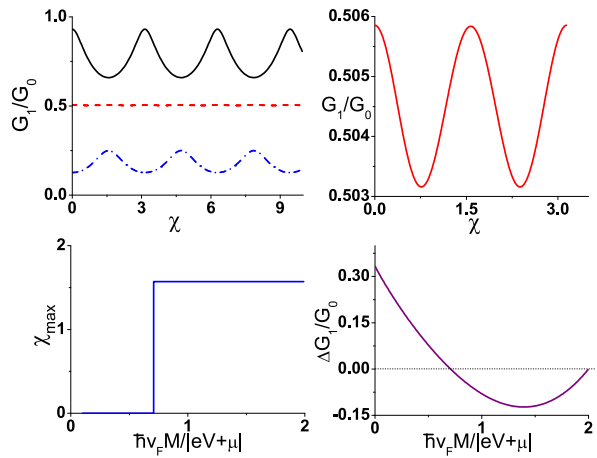}
\caption{Top left panel: Plot of $G_1/G_0$ versus $\chi$
for $\hbar v_F M/|eV+\mu|=0.1$ (black solid line), $0.7075$ (red
dashed line), and $1.4$ (blue dash-dotted line) for fixed $V$ and
$\mu$. Top right panel: Plot of $G_1/G_0$ versus $\chi$ at
$M=M^{\ast}$ showing the period halving. Bottom left panel: Plot of
$\chi_{\rm max}$ versus $\hbar v_F M/|eV+\mu|$ showing the
shift of conductance maxima. Bottom right panel: Plot of $\Delta G_1/G_0$
versus $\hbar v_F M/|eV+\mu|$ which crosses $0$ for $M < M_c$ at $M=
M^{\ast}$. The dotted line is a guide to the eye.} \label{fig5} \end{figure}

The conductance of such a junction is given by $G_1 = (G_0/2)
\int_{-\alpha_1}^{\alpha_2} T_1 \cos(\alpha) d \alpha$, where
$\alpha_{1,2}$ are determined from the solution of $\cos(\beta)=0$
for a given $M$ (Eq.\ (\ref{magwave1})). A plot of $G_1$ as a
function of $\hbar v_F M/|eV+\mu|$ and $\chi$ (for a fixed $eV$ and
$\mu$) is shown in Fig.\ \ref{fig4}. We find that the amplitude of
$G_1$ decreases monotonically as a function of $|M|$ reaching $0$ at
$M=M_c$ beyond which there are no propagating modes in region III.
Also, as we increase $M$, the conductance maxima shifts from $\chi=
n \pi$ to $ \chi= (n+1/2)\pi$ beyond a fixed value of $M^{\ast}(V)
\simeq \pm c_0 |eV+\mu|/(\hbar v_F)$ as shown in top left panel of
Fig.\ \ref{fig5}. Numerically, we find $c_0=0.7075$. At
$M=M^{\ast}$, $G_1(\chi=n\pi)=G_1(\chi=(n+1/2)\pi)$, leading to a
period halving of $G_1(\chi)$ from $\pi$ to $\pi/2$ . This is shown
in top right panel of Fig.\ \ref{fig5} where $G_1(M=M^{\ast})$ is
plotted as a function of $\chi$. We note that near $M^{\ast}$, the
amplitude of oscillation of $G_1$ as a function of $\chi$ becomes
very small so that $G_1$ is almost independent of $\chi$. In the
bottom left panel of Fig.\ \ref{fig5}, we plot $\chi=\chi_{\rm max}$
(the value of $\chi$ at which the first conductance maxima occurs)
as a function of $\hbar v_F M/|eV+\mu|$ which clearly demonstrates
the shift. This is further highlighted by plotting $\Delta G_1
=G_1(\chi=0)-G_1(\chi=\pi/2)$ as a function of $\hbar v_F
M/|eV+\mu|$ in the bottom right panel of Fig.\ \ref{fig5}. For
$M<M_c$, $\Delta G_1$ crosses zero at $M=M^{\ast}$ indicating the
position of the above-mentioned period halving. Thus we conclude
that the position of the conductance maxima depends crucially on
$\hbar v_F M/|eV+\mu|$ and can be tuned by changing either $M$ or
$V$.

The experimental verification of our results would involve
preparation of junctions by depositing ferromagnetic films on the
surface of a topological insulator. For the geometry shown in the
left panel of Fig.\ \ref{fig1}, we propose measurement of $G$ as a
function of $m_0$ whose magnitude and direction can be tuned by an
externally applied in-plane magnetic field for soft ferromagnetic
films \cite{tanaka1}. We predict that depending on $m_0$, $G$ should
demonstrate either a monotonically decreasing or an oscillatory
behavior as a function of $d$. Another, probably more experimentally
convenient, way to realize this effect would be to measure $V_c$ of
a junction of width $d$ for several $M$ and confirm that $V_c$
varies linearly with $M$ with a slope of $\hbar v_F/(2e)$, provided
$\mu$ and $d$ remain fixed. For the geometry depicted in the right
panel of Fig.\ \ref{fig1}, one would, in addition, need to create a
barrier by tuning the chemical potential of an intermediate thin
region of the sample as done earlier for graphene \cite{neto1}. Here
we propose measurement of $G_1$ as a function of $V_0$ (or
equivalently $\chi$) for several representative values of $m_0$ and
a fixed $V$. We predict that the maxima of the tunneling conductance
would shift from $\chi=n \pi$ to $\chi=(n+1/2)\pi$ beyond a critical
$m_0$ for a fixed $V$, or equivalently, below a critical $V$, for a
fixed $m_0$.

In conclusion, we have studied the transport of Dirac fermions on
the surface of a topological insulator in the presence of proximate
ferromagnetic films in two experimentally realizable geometries. Our
study unravels novel features of the junction conductances which
have no analog in either graphene or 2D Schr\"odinger electrons and
can be verified in realistic experimental setups.

\end{document}